\documentclass[final,times,5p]{elsarticle}
\usepackage{graphicx}
\usepackage{xspace}
\usepackage{amsmath}
\usepackage{amsfonts}
\usepackage{psfrag}
\usepackage[psamsfonts]{amssymb}
\usepackage[dvips]{color}
\usepackage{srcltx}
\usepackage{hyperref}

\newcommand{\TCM}[1]{\textcolor{black}{#1}}

\newcommand{\DD}{\ensuremath{D\overline{D}{}}\xspace}
\newcommand{\nDD}{non-\ensuremath{D\overline{D}{}}\xspace}

\newcommand{\PP}{\ensuremath{\psi(2S)}\xspace}
\newcommand{\JP}{\ensuremath{J/\psi}\xspace}
\newcommand{\DP}{\ensuremath{\psi(3770)}\xspace}
\newcommand{\psiDP}{$\psi(3770)$}

\newcommand{\BL}{\phantom{$\approx\,$}}

\newcommand{\ee}{\ensuremath{e^{+}e^{-}}\xspace}

\renewcommand{\epsilon}{\varepsilon}

\begin{document}
\begin{frontmatter}
  \title{
   Analysis of BaBar, Belle, BES-II, CLEO and KEDR data on 
   $\psi(3770)$ line shape and 
   determination of the resonance parameters
  }
\author[binp,nsu]{A.\,G.\,Shamov}
\author[binp,nsu]{K.\,Yu.\,Todyshev\corref{cor}}
\cortext[cor]{Corresponding author K.\,Yu.\,Todyshev, 
e-mail: todyshev@inp.nsk.su }
\address[binp]{Budker Institute of Nuclear Physics, Siberian Div., 
  Russian Acad. Sci., 630090, Novosibirsk, Russia}
\address[nsu]{Novosibirsk State University, 630090, Novosibirsk, Russia}
\begin{abstract}
  The available data on the \DD\ and inclusive hadronic
  cross sections in the \psiDP\ region from the BaBar, Belle, BES-II, 
  CLEO and KEDR
  experiments have been analyzed assuming that systematic uncertainties on
  cross sections measured by various detectors are not correlated.
  Four theoretical models predicting the $\psi(3770)$ line shape were
  considered for the \DD\ channel.
  All of them gave satisfactory description of the data. 
  The combined analysis of the \DD and inclusive hadronic
  channels was performed using the model
  based on the vector dominance approach and accounting for the contribution
  of the \PP tail to the \DD\ cross section. The following values of
  the mass, total width, electron width and decay probability to the 
  non-\DD\ states were obtained:
  \begin{center}
    \begin{tabular}{cccc}
      $M$ (MeV) & \BL\BL$\Gamma$ (MeV) & \BL\BL$\Gamma_{ee}$(eV)
      & \BL\BL$\mathcal{B}_{n\DD}$ \\*[0.5ex]
      3779.8$\pm$0.6 & \BL\BL 25.8$\pm$1.3  & \BL\BL 196$\pm$18 
      & \BL\BL 0.164$\pm$0.049
    \end{tabular}
  \end{center}
where the errors quoted include both statistical and dominant 
systematic uncertainties.
\end{abstract}
\end{frontmatter}
\section{Introduction}
\label{sec:intro}

Behaviour of the hadronic cross section in the vicinity of the \psiDP\ 
is an object of numerous 
studies, both experimental and theoretical, and a subject of many
discussions over the last ten years.
The main topics of the discussions are the resonance line shape and
the probability of decay to the non-\DD\ states. The data on the
\DD and inclusive hadronic cross section
expected from BES-III should shed light on both of them. While
waiting for these data, it would be useful to perform a joint analysis of 
all experimental data available so far, compare different
theoretical models related to the \psiDP\ line shape and estimate its
non-\DD\ decay probability. Some BES-III results on the \DD\ channel 
were already presented at the workshop on charm physics~\cite{FANG}
but have not yet been published.

A summary of the experimental data used in this study 
is presented in Table \ref{Tab:ResExp}. Five types of data were
involved: inclusive hadronic data in the form of the $R$ ratio
from BES and in the form of the detected cross section with a known 
detection efficiency from KEDR, the \DD\ cross section measured by 
BaBar and Belle 
with radiative return, the $\ee\!\to\!\DD(\gamma)$ cross section by
BES and CLEO, and, finally, the inclusive non-\DD cross section
extracted by BES using kaons of high momentum which
can not appear in decays of $D$-mesons not too far from the
production threshold.
The data for the $R$ measurement were collected by BES in December 2003,
but the data set collected in March 2003 and mentioned
in \cite{BES:DD2008Anomal} is, unfortunately, not available.

\renewcommand{\arraystretch}{1.05}
\begin{table}[h!]
\caption{\label{Tab:ResExp} Compilation of results near the  \psiDP. 
207 data points at $3.678 < W < 3.9$~GeV.}
\begin{center}
\begin{tabular}[c]{|l|l|} \hline Analysis &    Comments \\ \hline
 BES \cite{BES:R2006} & 68 points of $R(W)$ \\  
 inclusive hadrons               & 
\\ \hline
 BaBar  \cite{BABAR:DD2007} &  36 points, \\ 
 \DD          &     $\!\int\limits_{W\!-\!\Delta W/2}^{W\!+\!\Delta W/2}
 \!\!\!\!\!\!\sigma_{e^{+}e^{-}\to\DD}(W')\,dW'/\Delta W$,  \\  
          &   $\Delta W\!=\!5$~MeV\\  
 \hline
BES  \cite{BEStot2007}+\cite{BES:NonDD2008}
      &  1+1 point, $\sigma_{\ee\!\to \text{hadr}(\gamma)}(W)$, \\
inclusive hadrons, non-\DD  &  $\sigma_{\ee\to \text{non}\DD(\gamma)}(W)$, \\
                                  &                   $W\!=\!3773$ MeV \\\hline
 BES   \cite{BES:DD2008}  & 14 points, $\sigma_{\ee\to \DD(\gamma)}(W)$ \\  
 \DD    & \\  \hline
 Belle  \cite{BELLE:DD2008} &  9 points, \\
  \DD       &   $\int\limits_{W\!-\!\Delta W/2}^{W\!+\!\Delta W/2}
              \!\!\!\!\!\! \sigma_{e^{+}e^{-}\to\DD}(W')\,dW'/\Delta W$, \\
            &    $\Delta W\!=\!20$~MeV\\ \hline
CLEO   \cite{CLEO:DD}+\cite{Bonvicini:2013vxi} 
                       &  1+1 point, $\sigma_{\ee\!\to \text{hadr}(\gamma)}(W)$,  \\
inclusive hadrons, \DD  & $\sigma_{\ee\to \DD(\gamma)}(W)$,\\
& $W\!=\!3774\pm 1$ MeV \\\hline
KEDR \cite{KEDR2012Psi3770} &  17+21+38 points in 3 scans    \\
inclusive hadrons         &
              $\sigma^{\mathrm{visible}}_{\ee\!\to \text{hadr}(\gamma)}(W)$\\
\hline
\end{tabular}
\end{center}
\end{table}

\section{Inclusive hadronic cross section}

In the energy range from the \PP mass to the $\DD\pi$ threshold
the inclusive hadronic cross section can be parameterized as follows:
\begin{equation}
\begin{split}
 \sigma^{\mathrm{visible}}_{\ee\to \text{hadr}(\gamma)} = \: &
     (\epsilon_{\psi(2S)}\, \sigma^{\text{RC}}_{\psi(2S)} +\!
     \epsilon_{J/\psi}\, \sigma^{\text{RC}}_{J/\psi} +\!
     \epsilon_{\tau\tau}\,\sigma^{\text{RC}}_{\tau\tau} +
     \epsilon_{\text{uds}}\,\sigma^{\text{RC}}_{\text{uds}}) \,\:+\:\, \\
    &  ( \epsilon_{\DD}\,\sigma^{\text{RC}}_{\DD} +
      \epsilon_{n\DD}\,\mathcal{B}_{n\DD}\,
           \sigma^{\text{RC}}_{\DP}\,),    
\end{split}
\label{eq:SigMHobs}
\end{equation}
where  $\sigma^{\text{RC}}$'s are theoretical
cross sections, $\epsilon$'s are corresponding detection efficiencies.
The $\text{RC}$ superscript means that the cross
section has been corrected
for initial-state radiation (ISR) effects,
$n\DD$\, stands for the direct \DP\, decay to light
hadrons, the other (super/sub)scripts seem self-explanatory.
The detection efficiencies depend on energy weakly and monotonously.

In this work Eq.\eqref{eq:SigMHobs} was used for analysis of KEDR data in the
approximation of Ref.~\cite{KEDR2012Psi3770}.
Only the latter parenthesis in \eqref{eq:SigMHobs} is essential for 
\DP characteristics, the former one represents
the background which is subtracted by introducing additional parameters
in the fit.

The multihadron cross section
by BES was published in terms of $R$. That simplifies the background
subtraction but complicates a study of the \DD\, signal and inclusive
\nDD decays of \DP.
Indeed, the calculation
of $R$ requires weighting of different detection efficiencies entering
in \eqref{eq:SigMHobs} and the evaluation of ISR corrections for 
the \DD\, cross section, which
can not be done without assumptions about the $\sigma_{DD}(W)$ behaviour
and \DP parameters. 
To facilitate  this problem, 
$R$ was transformed to the 
excess of the observable cross section due to \DD production
and non-\DD decays of \DP: 
\begin{equation}
\delta\sigma_{\text{hadr}(\gamma)} =
   \sigma^{\text{RC}}_{\DD} +
      (\epsilon_{n\DD}/ \epsilon_{\DD})\:\mathcal{B}_{n\DD}\,
           \sigma^{\text{RC}}_{\DP}.   
\end{equation}
The required information
on the detection efficiency and on
the account for radiative corrections was taken from
Refs.~\cite{BES:R2006,Ablikim:2006zq}.
Some details on the transformation can be found
in the Appendix. Since the properties of the non-\DD\ decay 
are not known well, it was assumed that 
$\epsilon_{n\DD}/\!\epsilon_{\DD}\!= 1$.

\section{\DD\, cross section}

The \DD\ cross sections can be presented in the form
\begin{equation}
\sigma_{\DD}(W) =\frac{\pi\alpha^2}{3W^2}\,\, \beta_D^{3}\, 
   \left|F_D(W)\right|^2,  \:\:\:\: \beta_D = \sqrt{1-4m_D^2/W^2},
\label{eq:DD}
\end{equation}
where  $\beta_D$ is the meson velocity in the c.m. frame
and $F_D$ is a $D$ meson form factor. For determination
of the \DP\ parameters in the usual way it should be split into
independent parts
\begin{equation}
\label{eq:FF}
F_D(W) = F^{\psi(3770)}(W)\,e^{i\phi} + F^{\text{N.R.}}(W),
\end{equation}
where $F^{\psi}(W)$ is a P-wave Breit-Wigner amplitude
with energy-dependent total width $\Gamma(W)$, $F^{\text{N.R.}}(W)$ is
a non-resonant (with respect to \DP) part of the form factor
and $\phi$ is a relative phase shift.

In the spirit of the Vector
Dominance Model (VDM) one can assume that the form factor is saturated by
contributions of nearest vector mesons.
Below 3.9~GeV it can be reduced to
\begin{equation}
\label{eq:VDM}
F^{\text{N.R.}}(W) = F^{\psi(2S)}(W) + F_0,
\end{equation}
where $F_0$ is
a real constant representing contributions of $\psi(4040)$ and
higher $\psi$'s. 
Such an approach was used for determination of
\DP\, parameters in Ref.~\cite{KEDR2012Psi3770} and 
was employed in this work. 
In BES-III report~\cite{FANG} it is considered as one of two options.

Alternatively, the form factors following from the works
\cite{AchasovShestakov:2012,ChenZhao:2012,Cao:2014vca}
were implemented in the analysis procedure. In these cases the form factor
can not be split according to Eq.~\eqref{eq:FF}, thus the resonance
mass and widths can not be compared with those obtained in the VDM
approach.

The \DD\, cross sections published by BaBar and Belle can be directly
matched with the theoretical cross section in Eq.~\eqref{eq:DD}.
The $\DD(\gamma)$ cross sections published by BES and CLEO correspond to
$\sigma^{\text{RC}}_{\DD}(W)$ in Eq.~\eqref{eq:SigMHobs}: 
\begin{equation}
\sigma^{\text{RC}}_{\DD}(W) = \!\!\int\!\!
 \sigma_{\DD}\left(W^{\prime}\sqrt{1\!-\!x}\,\right)
   \mathcal{F}(x,W^{\prime\,2})\,G(W,W^{\prime})\,dW^{\prime}dx,
\label{eq:DDRC}
\end{equation}
where $\mathcal{F}(x,s)$ is the radiative correction kernel
depending on the fraction of $s$ lost
in the initial-state radiation \cite{KF} and
$G(W,W^{\prime})$ describes
a distribution of the total collision energy. The latter
is assumed to be Gaussian with the
standard deviation $\sigma_{W}$ which is about 1.3 and 2~MeV for BES and CLEO,
respectively~\cite{PDG:2016}.

\section{Data analysis}

A sum of likelihood functions for independent experiments is minimized with
\begin{equation}
 \mathcal{L}^{\text{exp}} = \mathcal{L}_{\text{data}}(f_N,\Delta_W)
+ \mathcal{L}_{\text{syst}}(f_N,\Delta_W),
\label{eq:LH}
\end{equation}
where $\mathcal{L}_{data}$ is the Poisson likelihood
function multiplied by two when the number
of observed events is known or just $\chi^2$
when only the cross section 
and its error are known. The integration
of the theoretical cross section to match BaBar and Belle 
data is performed numerically with an energy step of 0.5~MeV.

Expected values in the likelihood are calculated with the
additional free parameters $f_N$ and $\Delta_W$ 
specific for each experiment. They
account for systematic uncertainties of the experiment
in the cross section normalization
and in the energy scale, respectively.
Their variations are limited by the term
$\mathcal{L}_{\text{syst}}\!=\!(f_N\!-\!1)^2/\sigma_N^2\!+\!\Delta_W^2/\delta_W^2$
with the $\sigma_N$ and $\delta_W$ values
taken from the appropriate publication.
Thus, the uncertainty estimates returned by the fit reflect 
a statistical error as well as dominant systematic uncertainties.

The characteristic value of the normalization uncertainty $\sigma_N$
is about 10\% for the \DD\, cross section measurements and about 3\%
for the inclusive hadronic cross section. The typical energy uncertainty
is about 1-2~MeV for all experiments but KEDR, which
has a high-precision system of beam energy determination.

Due to the second term of Eq.~\eqref{eq:LH},
introduction of parameters $f_N$ and $\Delta_W$ does not reduce the
number of degrees of freedom. Up to nine additional free parameters
were used to subtract the background and exclude systematic
uncertainties related to calculations of the \PP radiative tail.
\TCM{The subtraction of the background in the KEDR data is described
in Ref.~\cite{KEDR2012Psi3770}. For the BES hadronic data
two parameters were introduced, $\Delta R_{uds}$ 
and $\Delta\Gamma_{ee}^{\PP}/\Gamma_{ee}^{\PP}$.
The former corrects the $uds$-background level, the latter gives the
relative correction to the value of the
\PP leptonic width used in Ref.~\cite{BES:R2006} for the radiative
correction calculation.}

\section{VDM fit results}

\begin{figure}[t]
\vspace{-16.5mm}
\includegraphics[width=1.04\columnwidth]{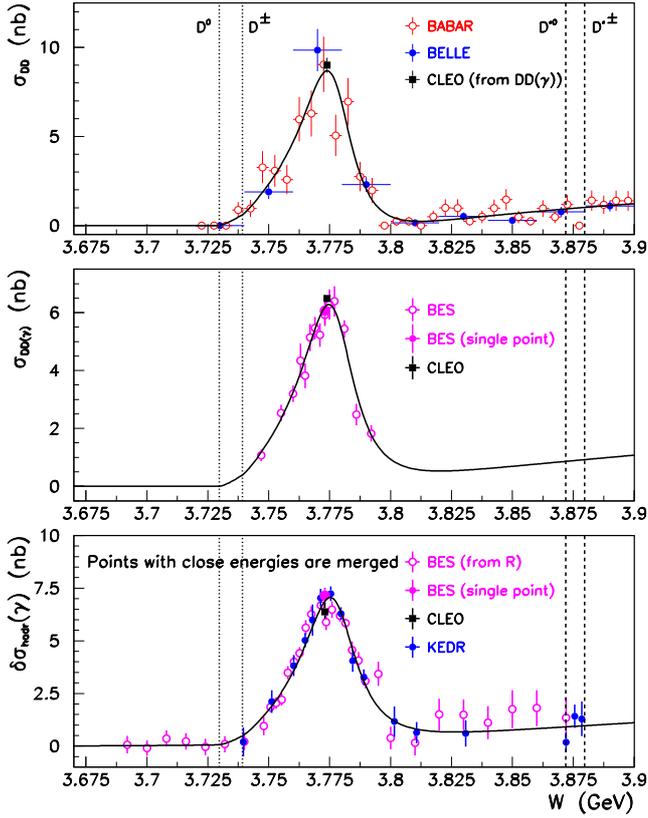}
\vspace{-9mm}
 \caption{
{\normalsize
The fit of data in the $\ee\!\to\!\DD$, $\ee\!\to\!\DD(\gamma)$ 
and $\ee\!\to\!hadrons(\gamma)$ channels.
For the latter only the excess of the cross section
associated with the \DD\, threshold is shown. The dotted and dashed lines mark
$D^0$, $D^{\pm}$ and $D^{*}$'s thresholds. \label{fig:fit}}}
\end{figure}

The results of the combined fit of all data in the VDM ansatz
are presented in Fig.~\ref{fig:fit}.  The fit
gives $\chi^2\!=\!230$ for 192 degrees of freedom which
corresponds to the $\chi^2$ probability of about 3.2\%.  Such a value
of $P(\chi^2)$ can be considered as satisfactory due to outliers
existing in some data sets. After a few points are removed, $P(\chi^2)$
reaches 17\%.

The fit gives the following values of the parameters
$f_N$ and $\Delta_W$ for \DD channels:
\begin{center}
\setlength{\tabcolsep}{10pt}
\begin{tabular}{l|c|c}
  {\it Experiment}  &  $f_N-1$ (\%)  & $\Delta_W$ (MeV)\\\hline
  Belle  &  -6.02 $\pm$ 6.55 & -1.23$\pm$1.16\\
  BaBar  &  0.17 $\pm$ 5.99 & 2.90$\pm$1.00\\
  BES    &  4.99 $\pm$ 3.97 & 0.40$\pm$0.19\\
\end{tabular}
\end{center}
They demonstrate that the experiments agree within the errors quoted.
The parameters $f_N$ were fixed at unity for
the data obtained by BES, CLEO and KEDR in the inclusive hadronic channel.
The corrections for the $R_{uds}$ and $\Gamma^{\PP}_{ee}$ values
used in the BES hadronic data analysis are small,
$\Delta R_{uds}=0.015\pm0.022$,\:
$\Delta\Gamma_{ee}^{\PP}/\Gamma_{ee}^{\PP}=0.022\pm0.014$.

As discussed in \cite{KEDR2012Psi3770}, for the form factor in Eq.~\eqref{eq:FF}
the likelihood function has two local minima with very close
values of $\chi^2$ at two values of the relative phase
$\phi$. This corresponds to two possible solutions for the \DP
parameters:\\[1pc]
\setlength{\tabcolsep}{3pt}
\begin{tabular}{cccccc}
  &  $M$ (MeV) &  $\Gamma$ (MeV) & $\mathcal{B}_{n\DD}$ & $\Gamma_{ee}$(eV)
                                                   & $\phi$ (deg) \\*[0.5ex]\hline
 1: & 3779.8$\pm$0.6 & 25.8$\pm$1.3 & 0.164$\pm$0.049 & 196$\pm$18 & 187$\pm$5\\
 2: & 3779.9$\pm$0.6 & 25.9$\pm$1.3 & 0.099$\pm$0.030 &328$\pm$18 & 227$\pm$3\\ 
\end{tabular}
\vspace*{1.ex}

The unitarity condition gives some arguments that the interference phase $\phi$
should be close either to zero or 180 degrees. For this reason the
first solution looks preferable.

The results obtained confirm the conclusions of the 
work~\cite{KEDR2012Psi3770} by
the KEDR collaboration. Namely,
the exotic assumptions are not necessary for the description of the 
\DD\ line shape and the value of the \DP\ mass
is almost 7~MeV higher than the result of the PDG fit~\cite{PDG:2016}
performed using measurements which do not account for effects of the
resonance-continuum interference.
Similar results
on the mass and total width were presented by BES-III~\cite{FANG}:
$M\!=\!3781.5\pm0.3$, $\Gamma\!=\!25.2\pm0.7$~MeV (statistical errors only)
although the value of the relative phase differs: 
$\phi\!=\!208\pm4$~degrees.
It should be noted that in~\cite{FANG} it was
not specified which of possible solutions was chosen. 

The fit gives a large probability of the non-\DD decays 
$0.164\pm0.049$ in agreement with the results by 
the BES collaboration
obtained using 'inclusive non-\DD selection'~\cite{BES:NonDD2008}:
$$\mathcal{B}_{n\DD}=0.151\pm0.056\pm0.018.$$
The formal statistical significance of the result is about 3.3$\,\sigma$, 
however, the total error of 0.049 is dominated by systematic uncertainties 
in the \DD\ normalization scale and can be underestimated. The exclusion of 
BES data on $\ee\!\to\!\DD(\gamma)$~\cite{BES:DD2008} reduces the
result to $\ 0.126\pm 0.056$ (2.2$\,\sigma$).

The results presented above were obtained without the correction of
the $R$ values by BES to the detection efficiency dependence on the
assumption on \DP shape and parameters which were used in the
analysis~\cite{BES:R2006}.
To check the corresponding systematic uncertainties this correction
was applied as described in the Appendix. The following variations of
the main \DP parameters were obtained: $\delta M= -0.3$~MeV,
$\delta\Gamma=0.4$~MeV, $\delta\Gamma_{ee}=4.2$~eV,
$\delta\mathcal{B}_{n\DD}=0.007$.  They are not large compared to the
total errors of corresponding parameters.

\section{Alternative \DD cross section models}

Recently a few theoretical works appeared that discussed the \DP line shape
in the \DD decay channel, their comprehensive list can be found
elsewhere~\cite{Du:2016qcr}.
Here we consider three works:
N.~N.~Achasov and G.~N.~Shestakov \cite{AchasovShestakov:2012} suggested
a model of the $D$ meson form factor which meets the elastic
unitarity requirement; G.-Y.~Chen and Q. Zhao \cite{ChenZhao:2012}
studied the line shape of the \DD\ cross section within an effective
field theory, while X.~Cao and H.~Lenske \cite{Cao:2014vca} accounted
for the interactions between \DP\ and the \DD\
continuum in the approach suggested by U.~Fano~\cite{Fano}.
The number of free parameters in these models is equal to six.

In all three cases the predictions for the inclusive
hadronic cross section for the non-negligible $\mathcal{B}_{n\DD}$
are absent.
For this reason the models were employed to fit 
the \DD\ cross section measured by BaBar, Belle, BES and CLEO and compare
the fit quality with that for the VDM inspired model. The following 
values of $\chi^2$ for the fits in the two energy regions were obtained:
\begin{center}
\begin{tabular}{c|rc|rr}
  {\it Model}\BL & $\chi^2/N_{DoF}$ & $P(\chi^2)$\%
                   & $\chi^2/N_{DoF}$ & $P(\chi^2)$\%\\*[0.25ex]
\hline
VDM (this work)  & 68.1 / 54 & 9.4 & 40.4 / 34 & 20.8\\
A-S \cite{AchasovShestakov:2012} & 75.4 / 54 & 2.9 & 43.9 / 34 & 11.9\\
C-Z  \cite{ChenZhao:2012} & 75.6 / 54 & 2.8  &  43.9 / 34 & 11.9\\
C-L  \cite{Cao:2014vca} & 75.5 / 54 & 2.8  &  44.9 / 34 & 10.0\\
\hline
 $W$& \multicolumn{2}{|c}{$<3.9$ GeV} & \multicolumn{2}{|c}{$<3.82$ GeV} \\
\end{tabular}
\end{center}
The alternative models do not improve the description
of available \DD\, data.

\begin{figure}[t]
\vspace{-7.mm}
\includegraphics[width=1.04\columnwidth]{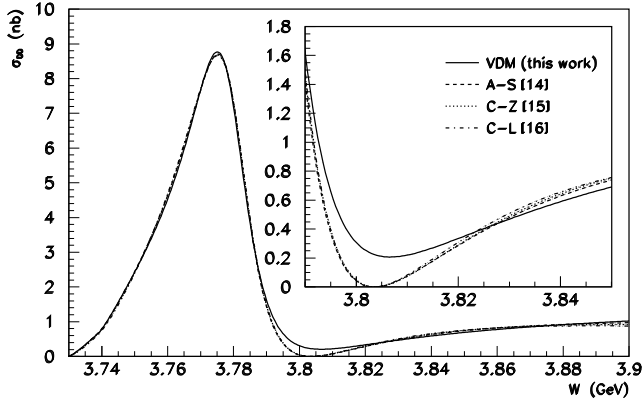}
\vspace{-9mm}
 \caption{{\normalsize
Comparison of \DP\ line shapes for the models considered.
\label{fig:cmp}}}
\end{figure}

The line shapes with parameters of models obtained in the fit are
presented in Fig.~\ref{fig:cmp}. The resulting line shapes for 
the models~\cite{AchasovShestakov:2012,ChenZhao:2012,Cao:2014vca} 
are surprisingly similar.
All of them predict a zero \DD\ cross section slightly above 3.8~GeV. 
The cross section can not drop to zero in the model in which the amplitude 
is a sum of Breit-Wigner shapes.
The value of $\chi^2$ for six data points in the energy range 
from 3.795 to 3.82~GeV is about 5.86 for the VDM assumptions and 
about 8.1 for the alternative models,
thus vanishing of the cross section can not be either excluded or confirmed.

\section{Conclusion}
A joint analysis of all data on cross sections 
for hadron production in $e^+e^-$ annihilation around the $\psi(3770)$
resonance published so far leads, in our opinion, to the following conclusions:
\begin{itemize}
\item
Account of the resonance-continuum interference
in any reasonable way solves the problem of the \DP\ line shape. The
existing data do not allow a selection of the best model among 
those considered.
\item
The \DP\ parameters obtained ignoring the resonance-continuum interference
are not accurate and should not be used. In particular, it concerns
various results on the
difference of the \DP\ and \PP\ masses presented in Ref.~\cite{PDG:2016}.
\item
The \DP\ cross section measured in the \DD\ channel is less than the total
one by a factor of $0.836\pm0.049$, thus the problem
of the non-\DD\ decays is still pending.
\end{itemize}
The data on the \DD\ cross section and the inclusive
hadronic cross section from BES-III are eagerly awaited.

\section*{Acknowledgments}

The authors are grateful to V.~E.~Blinov and S.~I.~Eidelman
for supporting the work and stimulating discussions.

\section*{Appendix}

The $R$ data by BES was transformed to the cross section as follows:
\begin{equation}
\delta\sigma_{\text{hadr}(\gamma)}(W) =
    F^{\text{RC}}(W)\,
       \left( R\: \epsilon_h(W)/\epsilon^{\prime}_h(W)
              - R_{uds} \right)\,
                           \sigma^{\text{B}}_{\mu\mu}(W),
\label{eq:dsig}
\end{equation}
where $\sigma^{\text{B}}_{\mu\mu}(W)=4\pi\alpha^2/3W^2$
is the Born cross section of the
dimuon production, $R_{uds}=2.141$ is the light quark contribution in
$R$ according to~\cite{BES:R2006},
$\epsilon_h(W)$ is the net hadronic detection efficiency used for the $R$
calculation in Ref.~\cite{BES:R2006},
$\epsilon^{\prime}_h(W)$ is the hadronic detection efficiency
reweighted for the current assumption on the \DP shape and parameters.

The radiative correction factor $F^{\text{RC}}(W)$ for \DP production was
calculated as in Ref.~\cite{BES:R2006} using the values of 
the mass $M=3772.2$~MeV and the total width
$\Gamma = 26.9$~MeV~\cite{Ablikim:2006zq}.

The correction for the detection efficiency variation 
was applied iteratively. In terms of Eq.~\eqref{eq:SigMHobs}
the net detection efficiency for hadronic
events can be defined as follows:
\begin{equation}
\epsilon_h=\sigma^{\mathrm{visible}}_{\ee\to \text{hadr}(\gamma)}/
\sigma^{\text{RC}}_{\text{hadr}},
\label{eq:eps}
\end{equation}
where $\sigma^{\text{RC}}_{\text{hadr}}$ is the sum of hadronic cross
sections.
The $\epsilon_h(W)$ dependence
presented in Fig.~2 of Ref.~\cite{BES:R2006} was fitted
to extract the values of $\epsilon_{\text{uds}}$,
$\epsilon_{\JP}$, $\epsilon_{\PP}$ and
$\epsilon_{\DD}$ entering in Eq.~\eqref{eq:SigMHobs}. The
contribution of the $\tau$ pair production was neglected.
 The parameters and assumptions required for the calculation of 
$\sigma^{\text{RC}}_{\text{hadr}}$
were the same as in Ref.~\cite{BES:R2006}. At the first iteration the
fit of cross sections 
was performed with $\epsilon_h/\epsilon^{\prime}_h=1$ which
allowed to calculate the net efficiency $\epsilon^{\prime}_h(W)$
for the second iteration using Eqs.~\eqref{eq:eps} and
\eqref{eq:SigMHobs}.

At the first approximation the values of 
$\delta\sigma_{\text{hadr}(\gamma)}$ obtained as described above
do not depend on assumptions on the \DP shape and parameters.
Since $\epsilon_h(W)$ was taken from the plot, the correction for the
efficiency variation is not accurate. We used it only for the check
of the systematic uncertainty.

\end{document}